\documentclass[12pt]{article}
\usepackage{epsf}

\begin{document}

\newcommand{\gtrsim}{\mathop{}_{\textstyle \sim}^{\textstyle >}}
\newcommand{\lesssim}{\mathop{}_{\textstyle \sim}^{\textstyle <} }

\newcommand{\rem}[1]{{\bf #1}}

\renewcommand{\thefootnote}{\fnsymbol{footnote}}
\setcounter{footnote}{0}
\begin{titlepage}

\def\thefootnote{\fnsymbol{footnote}}

\begin{center}

\hfill TU-821\\
\hfill July, 2008\\

\vskip .75in

{\Large \bf 
Long-Lived Unstable Superparticles at the LHC
}

\vskip .75in

{\large
Koji Ishiwata, Takumi Ito and Takeo Moroi
}

\vskip 0.25in

{\em
Department of Physics, Tohoku University,
Sendai 980-8578, JAPAN}

\end{center}
\vskip .5in

\begin{abstract}

  In various models of supersymmetry (SUSY), the lightest
  superparticle in the minimal SUSY standard model sector, which we
  call MSSM-LSP, becomes unstable.  Then, we may observe the decay of
  the MSSM-LSP in the detector at the LHC experiment.  We show that
  the discovery of such a decay process (and the determination of the
  lifetime of the MSSM-LSP) may be possible at the LHC even if the
  decay length of the MSSM-LSP is much longer than the size of the
  detector; sizable number of the MSSM-LSPs decay inside the detector
  if the lifetime is shorter than $10^{-(3-5)}\ {\rm sec}$.  We also
  discuss the implications of the study of the MSSM-LSP decay for
  several well-motivated SUSY models.

\end{abstract}

\end{titlepage}

\renewcommand{\thepage}{\arabic{page}}
\setcounter{page}{1}
\renewcommand{\thefootnote}{\#\arabic{footnote}}
\setcounter{footnote}{0}

Supersymmetry (SUSY) is a well-motivated target of the LHC experiment.
Indeed, not only the discovery but also detailed studies of the
superparticles are possible at the LHC experiment if superparticles
are within the kinematical reach \cite{AtlasTdr,CmsTdr}.  In many of
the studies, it is assumed that $R$-parity is conserved, and that the
lightest superparticle (LSP) is the lightest neutralino.  If so, all
the produced superparticles cascade down to the lightest neutralino
just after the production and SUSY events are characterized by large
missing $p_T$.

Even though the lightest superparticle in the
minimal-SUSY-standard-model (MSSM) sector, which we call MSSM-LSP, is
often assumed to be the lightest neutralino and is stable, it is not
always the case.  Various scenarios where the MSSM-LSP becomes
unstable have been proposed.  One important example is the
gauge-mediated SUSY breaking scenario \cite{GMSB}, where the SUSY
breaking in the MSSM sector is mediated via the standard-model gauge
interaction so that SUSY-induced flavor violation are strongly
suppressed.  In the gauge-mediated model, the gravitino becomes the
LSP and the MSSM-LSP decays into the gravitino.

Another example is $R$-parity violation, with which the MSSM-LSP becomes
unstable.  Usually, $R$-parity conservation is assumed to realize the
LSP dark matter scenario.  However, LSP can be dark matter even if the
$R$-parity is violated; if the $R$-parity violation is weak enough, the
lifetime of the LSP becomes longer than the present age of the universe.
This possibility becomes important when the gravitino is the LSP
\cite{Takayama:2000uz,Buchmuller:2007ui}, because it has several
advantages.  In such a case, the thermal leptogenesis
\cite{Fukugita:1986hr}, which requires relatively high reheating
temperature \cite{TR_Leptogenesis}, may be possible without conflicting
with the constraints from big-bang nucleosynthesis \cite{GravitinoBBN}
and the overproduction of the gravitino \cite{Moroi:1993mb}.  In
addition, if the gravitino is dark matter with $R$-parity violation, a
fraction of gravitino dark matter decays until the present epoch.  The
decay becomes a source of the high energy cosmic rays.  In particular,
recently, it has been shown that the anomalous excesses of the
$\gamma$-ray and positron fluxes observed by EGRET \cite{egret} and HEAT
\cite{heat} experiments, respectively, can be simultaneously explained
in the gravitino dark matter scenario with $R$-parity violation if the
lifetime of the gravitino is about $10^{26}\ {\rm sec}$
\cite{Ibarra:2008qg,Ishiwata:2008cu}.  In such a scenario, the MSSM-LSP
decays mainly via the $R$-parity violating interaction with the lifetime
of $10^{-(5-6)}\ {\rm sec}$.  Discovery of the MSSM-LSP with such a
lifetime may give us a hint to understand the origin of high energy
cosmic rays.

The experimental search for the decay of the MSSM-LSP gives important
test of the scenarios with unstable MSSM-LSP.  Since the
superparticles are expected to be copiously produced at the LHC, we
may have a chance to find the signal of the decay of MSSM-LSP.  With
long-lived unstable MSSM-LSP, we may see the decay at the LHC
experiment in the form of a displaced vertex from the interaction
point, non-pointing particle, and/or a disappearance of high $p_T$
tracks.  Discovery of the MSSM-LSP decay is very important to
understand the property of the MSSM-LSP.  However, if the decay length
of the MSSM-LSP is much longer than the size of the detector, most of
the MSSM-LSPs escape from the detector before they decay.  In such a
case, the typical signal of the SUSY events are the same as the case
where the MSSM-LSP is stable, and the discovery of the decay becomes
statistically non-trivial.

In this letter, we discuss the possibility of discovering the decay of
the MSSM-LSP at the LHC experiment, paying particular attention to the
case where the decay length is much longer than the size of the
detector.  We will show that the discovery may be possible if the
lifetime is shorter than $\tau\lesssim 10^{-(3-5)}\ {\rm sec}$.  Then,
we consider the implication of the result for several types of models
with unstable MSSM-LSP.  Furthermore, if the decay is observed, the
lifetime of the MSSM-LSP can be constrained
\cite{Ambrosanio:2000ik,Kawagoe:2003jv}; we also discuss the
possibility of determining the lifetime of long-lived MSSM-LSP.

Let us start with discussing the basic formulae.  If the MSSM-LSP is
unstable, a fraction of the MSSM-LSPs produced at the LHC experiment
decay inside the detector.  If the MSSM-LSP (with its lifetime $\tau$)
has the velocity $v$, the decay probability before propagating
the distance $L$ is given by $P_{\rm dec}(L)=1-e^{-L/v\gamma\tau}$,
where $\gamma=(1-v^2/c^2)^{-1/2}$ (with $c\simeq 3.0\times 10^8\ {\rm
  m/sec}$ being the speed of light).  Then, denoting the
pseudo-rapidity of the MSSM-LSP as $\eta\equiv -\ln\tan(\theta/2)$
(with $\theta$ being the angle from the beam axis), the number of the
MSSM-LSPs which decay inside the detector is given by
\begin{eqnarray}
  N_{\rm dec} = N_{\rm tot} \int d\eta dv
  f(\eta,v) (1 - e^{-l^{\rm (max)}(\eta)/v\gamma\tau}),
\end{eqnarray}
where $N_{\rm tot}$ is the total number of MSSM-LSP, $l^{\rm
  (max)}(\eta)$ is the distance to the outer boundary of the detector
from the interaction point, and $f(\eta,v)\equiv N_{\rm tot}^{-1}
dN/d\eta dv$ is the distribution function of the MSSM-LSP; $\int d\eta
dv f(\eta,v)=1$.

In our following discussion, it is convenient to define
\begin{eqnarray}
  L^{\rm (eff)} \equiv 
  \int d\eta dv f(\eta,v) \frac{c l^{\rm (max)}(\eta)}{v\gamma}.
  \label{Leff}
\end{eqnarray}
Then, when the size of the detector is much smaller than the decay
length $c\tau$, we obtain
\begin{eqnarray}
  N_{\rm dec} = N_{\rm tot} \frac{L^{\rm (eff)}}{c\tau}.
  \label{N_dec}
\end{eqnarray}
When $c\tau\lesssim N_{\rm tot} L^{\rm (eff)}$, we expect several
decay events inside the detector.  Since, typically, $v\sim c$ and
$\gamma\sim O(1-10)$, $L^{\rm (eff)}$ becomes comparable to the size
of the detector, as we will see in the following.  Thus, roughly
speaking, the number of the decay inside the detector is determined by
the size of the detector and the total cross section for the SUSY
events.  In our analysis, for simplicity, we approximate the shape of
the detector as a cylinder with the radius $l_T^{\rm (max)}$ and the
half-length (to the $z$-direction) $l_z^{\rm (max)}\equiv l_T^{\rm
  (max)}/\tan\theta_{\rm edge}$:
\begin{eqnarray}
  l^{\rm (max)} (\eta) = 
  \left\{ \begin{array}{ll}
      l_T^{\rm (max)} / \sin\theta & ~:~ \eta < \eta_{\rm edge}
      \\
      l_z^{\rm (max)} / |\cos\theta| & ~:~ \eta > \eta_{\rm edge}
    \end{array} \right. ,
  \label{lmax}
\end{eqnarray}
where $\eta_{\rm edge}=-\ln\tan(\theta_{\rm edge}/2)$.  From the muon
chamber layout of the ATLAS detector, in our Monte Carlo (MC)
analysis, we take \cite{AtlasTdr}\footnote
{The end-cap of the ATLAS detector covers only up to $|\eta|<2.7$
  \cite{AtlasTdr}.  We have checked that most of the MSSM-LSPs are
  within this region, and hence we do not impose a cut on $\eta$ for
  simplicity.}
\begin{eqnarray}
  l_T^{\rm (max)}=10\ {\rm m}, ~~~\eta_{\rm edge}=1.0.
  \label{boundary}
\end{eqnarray}

Now, we calculate how many MSSM-LSPs decay inside the detector.  The
details depend on the MSSM parameters, and on what the MSSM-LSP is.
When the MSSM-LSP is unstable, charged (or even colored) MSSM-LSP is
phenomenologically viable.  In the following, we discuss two of the
important cases; one is the case where the lightest neutralino
$\chi^0_1$ is the MSSM-LSP while in the other case, the lighter stau
$\tilde{\tau}$ is the MSSM-LSP.  For example, in the gauge-mediated
model, they are two of the important candidates of the MSSM-LSP.

\begin{figure}[t]
 \centerline{\epsfxsize=0.55\textwidth\epsfbox{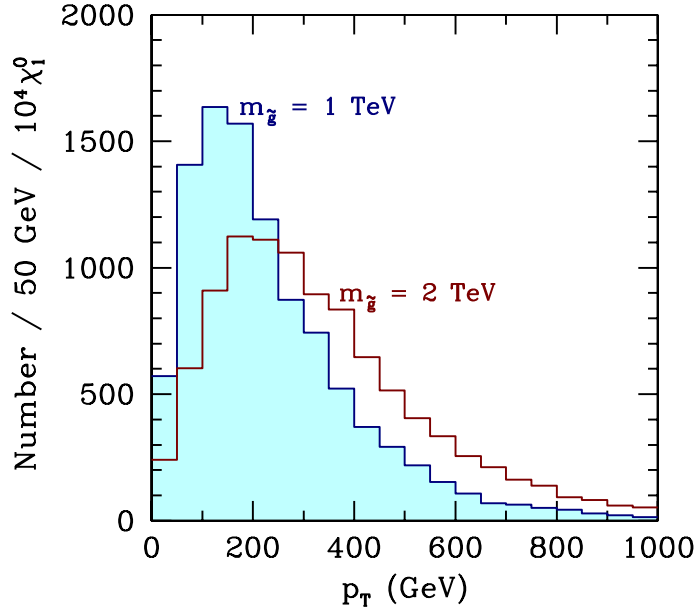}}
 \caption{\small Distribution of $p_T$ of the final-state lightest
   neutralino with $10^4$ samples for the $\chi^0_1$-MSSM-LSP case.
   We take $m_{\tilde{g}}=1\ {\rm TeV}$ (shaded) and $2\ {\rm TeV}$
   (unshaded).}
\label{fig:histptbino}
\end{figure}

First, we consider the case where the lightest neutralino $\chi^0_1$
is the MSSM-LSP, and is a long-lived unstable particle.  Even though
$\chi^0_1$ is invisible, spectacular signal may arise if $\chi^0_1$
decays inside the detector.  For example, in the models mentioned
above (i.e., the gauge-mediated model or the model with $R$-parity
violation), $\chi^0_1$ decays into $\gamma$ or $Z$ boson and an
invisible particle. If we can find a production of high energy
$\gamma$ or the decay products of $Z$ from the point which is
displaced from the interaction point, like non-pointing photon
\cite{Kawagoe:2003jv}, it can be identified as the signal of the decay
of $\chi^0_1$.  (The study of these signals may require that the decay
occurs at an inner region of the detector, which reduces the fiducial
volume compared to Eqs.\ (\ref{lmax}) and (\ref{boundary}).  For more
details, see later discussion.)  Signal of those events should be
searched at the off-line analysis.  In order to record those events,
trigger may be an issue.  One possibility is to use the missing $p_T$
trigger.  Because two $\chi^0_1$s are produced in the SUSY events,
even if one of $\chi^0_1$ decays inside the detector to be the signal
event, the other $\chi^0_1$ is expected to escape from the detector.
(Notice that we consider the case that $c\tau$ is much longer than the
size of the detector.)  Such a non-decaying $\chi^0_1$ should be a
source of large missing $p_T$.  With the MC analysis, we calculate the
distribution of the $p_T$ of $\chi^0_1$.  The distribution for
$m_{\tilde{g}}=1\ {\rm TeV}$ and $2\ {\rm TeV}$ are shown in Fig.\
\ref{fig:histptbino}.  With the present choice of parameters, more
than $80-90\ \%$ of $\chi^0_1$s have $p_T$ larger than $100\ {\rm
  GeV}$ for $m_{\tilde{g}}=1-2\ {\rm TeV}$.  Thus, assuming that the
escaping $\chi^0_1$ is the dominant source of the missing $p_T$, most
of the signal events have large missing $p_T$.  Thus, if a relevant
missing $p_T$ trigger is implemented, the event can be recorded.

In order to estimate how many $\chi^0_1$s decay inside the detector,
we perform MC analysis.  At the LHC experiment, MSSM-LSP is mostly
from the productions of colored superparticles: $pp\rightarrow
\tilde{g}\tilde{g}$, $\tilde{g}\tilde{q}$, and $\tilde{q}\tilde{q}'$
(with $\tilde{g}$ and $\tilde{q}$ being the gluino and squark,
respectively).  The cross sections for these processes as well as the
decay chains of the superparticles depend on the MSSM parameters.
Here, as a well-motivated example, we adopt the gauge-mediated model
to calculate these quantities.  (Notice that, even though we use the
gauge-mediated model to fix the underlying parameters, it is just for
simplicity.)  In our study, the SUSY events are generated as follows:
\begin{itemize}
\item[1.] Mass spectrum of the superparticles and their decay rates
  are calculated in the framework of the gauge-mediated model.  Here,
  the simplest gauge-mediated model is adopted where the model is
  parametrized by $\tan\beta$ (i.e., the ratio of the vacuum
  expectation values (VEVs) of up- and down-type Higgs bosons), number
  of vector-like messenger multiplets (in ${\bf 5}+{\bf \bar{5}}$
  representation of grand-unified $SU(5)$ group) $N_{\bf 5}$, the
  messenger scale $M_{\rm mess}$, and $\Lambda$ which is the ratio of
  the $F$-component of the SUSY breaking field to its VEV.
\item[2.] SUSY events expected at the LHC experiment are generated
  with $\sqrt{s}=14\ {\rm TeV}$.
\item[3.] In each event, decay chains of superparticles are followed
  and the resultant momentum distribution of the MSSM-LSP is obtained.
  Then, $L^{\rm (eff)}$ is calculated.
\item[4.] From the event samples, we calculate the number of the
  MSSM-LSP decay inside the detector as a function of $\tau$.
\end{itemize}
In our analysis, we use ISAJET package \cite{Paige:2003mg} for the
first step while, for second and third steps, HERWIG package
\cite{Herwig} is used.

\begin{figure}[t]
  \centerline{\epsfxsize=0.55\textwidth\epsfbox{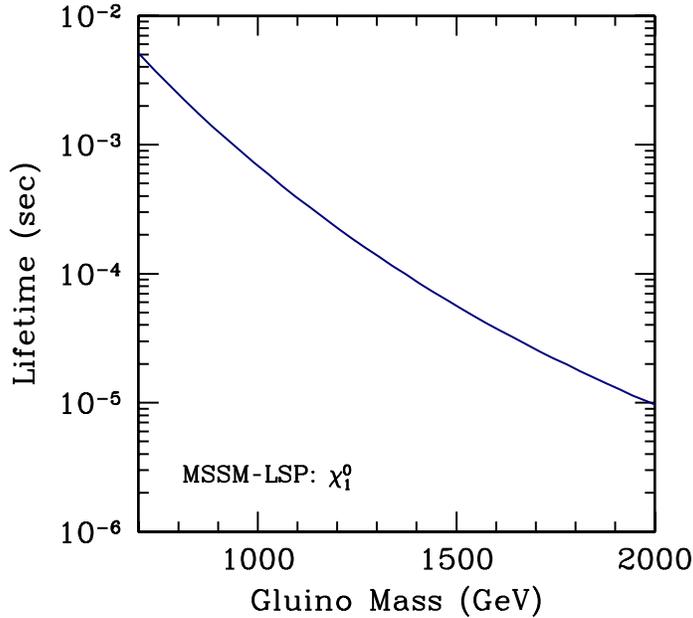}}
  \caption{\small The lifetime of $\chi^0_1$ which gives $N_{\rm
    dec}=10$ (with ${\cal L}=100\ {\rm fb}^{-1}$) as a function of the
    gluino mass.}
\label{fig:taumax_bino}
\end{figure}

For the study of the $\chi^0_1$-MSSM-LSP case, we take $\tan\beta=20$,
$N_{\bf 5}=1$, $M_{\rm mess}=10^7\ {\rm GeV}$, and $\Lambda$ is fixed
to determine the mass scale of MSSM particles (in particular, in the
present analysis, the gluino mass).  With the above choice of
parameters, the MSSM-LSP is Bino-like lightest neutralino, and its
mass is given by $m_{\chi^0_1}=175\ {\rm GeV}$, $360\ {\rm GeV}$, and
$500\ {\rm GeV}$ for the gluino mass $m_{\tilde{g}}=1\ {\rm TeV}$,
$1.5\ {\rm TeV}$, and $2\ {\rm TeV}$, respectively.\footnote
{With the present choice of parameters, the Higgs mass becomes smaller
than the present experimental bound of $114.4\ {\rm GeV}$
\cite{Yao:2006px} when $m_{\tilde{g}}\lesssim 950\ {\rm GeV}$.
Because we choose the gauge-mediated model just as an example of the
SUSY model to fix the mass spectrum of MSSM particles, and also
because the Higgs mass is sensitive to the masses of stops, we do not
take the Higgs-mass constraint so seriously and extend our study to
the parameter region of $m_{\tilde{g}}\lesssim 950\ {\rm GeV}$.  If
one is interested in the case of the gauge-mediated model, only the
results for $m_{\tilde{g}}\gtrsim 950\ {\rm GeV}$ are relevant.}

With the MC analysis, we found that $L^{\rm (eff)}$ is not sensitive
to the mass spectrum of superparticles, and is $\sim 10\ {\rm m}$;
$L^{\rm (eff)}=8.5\ {\rm m}$, $9.9\ {\rm m}$, and $11.4\ {\rm m}$ for
$m_{\tilde{g}}=1\ {\rm TeV}$, $1.5\ {\rm TeV}$, and $2\ {\rm TeV}$,
respectively.  We can see that $L^{\rm (eff)}$ is slightly enhanced as
$m_{\tilde{g}}$ increases, which is due to the decrease of the
averaged velocity of the MSSM-LSP for larger value of $m_{\tilde{g}}$.
On the contrary, the total cross section for the SUSY events strongly
depends on the masses of superparticles; the total cross section is
given by $1240\ {\rm fb}$, $80\ {\rm fb}$, and $10\ {\rm fb}$, for
$m_{\tilde{g}}=1\ {\rm TeV}$, $1.5\ {\rm GeV}$, and $2\ {\rm TeV}$,
respectively.

In Fig.\ \ref{fig:taumax_bino}, we plot the value of $\tau$ which
gives $N_{\rm dec}=10$ (with ${\cal L}=100\ {\rm fb}^{-1}$) as a
function of the gluino mass. (Since $N_{\rm dec}\propto\tau^{-1}$, the
lifetime which gives a different value of $N_{\rm dec}$ can be
calculated from the figure.)  For $m_{\tilde{g}}=1\ {\rm TeV}$, $1.5\
{\rm TeV}$, and $2\ {\rm TeV}$, $N_{\rm dec}\geq 10$ requires $\tau$
to be smaller than $7\times 10^{-4}\ {\rm sec}$, $6\times 10^{-5}\
{\rm sec}$, and $1\times 10^{-5}\ {\rm sec}$, respectively.  Thus,
when the lifetime is shorter than $10^{-(3-5)}\ {\rm sec}$, the number
of the decay of the MSSM-LSP inside the detector turns out to be
larger than $\sim O(1)$.

\begin{figure}[t]
  \centerline{\epsfxsize=0.55\textwidth\epsfbox{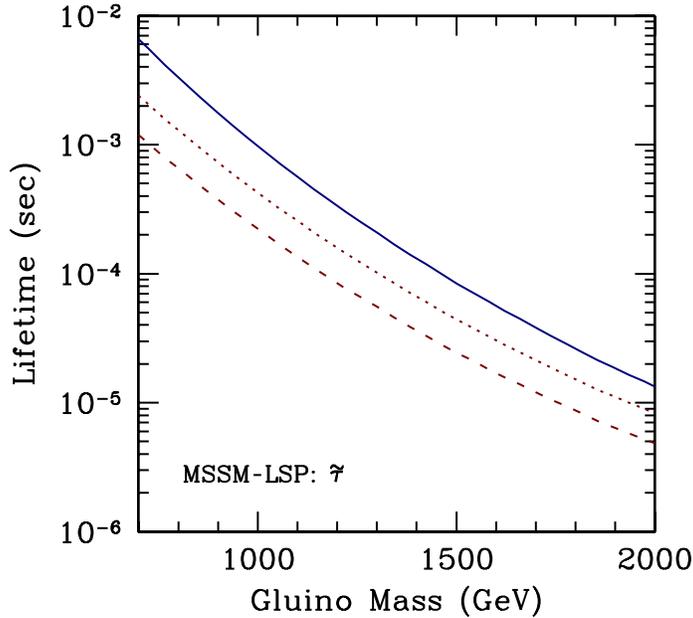}}
  \caption{\small The lifetime of $\tilde{\tau}$ which gives $N_{\rm
    dec}=10$ as a function of the gluino mass (solid), and that which
    corresponds to $N_{\rm dec}^{(v'<v_{\rm max})}=10$ with $v_{\rm
    max}=0.9$ (dotted) and $0.8$ (dashed).  The integrated luminosity
    is taken to be ${\cal L}=100\ {\rm fb}^{-1}$.}
\label{fig:taumax_stau}
\end{figure}

Next, we consider the decay of $\tilde{\tau}$-MSSM-LSP.  We expect
several possibilities to find the signal of the decay of
$\tilde{\tau}$.  In the models we introduced, $\tilde{\tau}$ decays
into tau lepton and an invisible particle.  For example, in the
gauge-mediated model, the invisible particle is the gravitino while,
in the scenario with the $R$-parity violation, it is a neutrino.  In
those cases, we may find a displaced decay vertex by observing decay
product(s) of the tau lepton.\footnote
{Otherwise, if selectron (smuon) is the MSSM-LSP, it decays into
  electron (muon) and an invisible particle.  Then, energetic charged
  tracks from displaced vertices are the target.}
In addition, if $\tilde{\tau}$ propagates $O(10\ {\rm cm})$ or so,
hits in inner trackers should exist.  Then, if we do not observe
corresponding hits in calorimeters or in muon detector, we may
identify such a short track as a signal of the $\tilde{\tau}$ decay
inside the detector.  Furthermore, using the fact that two staus are
produced, we can simply look for events with only one $\tilde{\tau}$
track (which may be useful to count the number of staus which decay
very inside the detector for the determination of $\tau$).  For the
case of $\tilde{\tau}$-MSSM-LSP, the trigger may not be an issue
because at least one high $p_T$ charged track (i.e., $\tilde{\tau}$)
exists in the SUSY event; then, we expect that we can use the muon
trigger.

Here and hereafter, in the study of the case where $\tilde{\tau}$ is
the MSSM-LSP, we adopt the underlying parameters of $\tan\beta=20$,
$N_{\bf 5}=2$, and $M_{\rm mess}=10^7\ {\rm GeV}$, with which
$\tilde{\tau}$ becomes the MSSM-LSP.  With such a choice of
parameters, the mass of $\tilde{\tau}$ becomes larger than $100\ {\rm
  GeV}$ when $m_{\tilde{g}}\gtrsim 650\ {\rm GeV}$, and is $170\ {\rm
  GeV}$, $270\ {\rm GeV}$, and $370\ {\rm GeV}$, for $m_{\tilde{g}}=1\
{\rm TeV}$, $1.5\ {\rm TeV}$, and $2\ {\rm TeV}$, respectively.  We
follow the same procedure as the case of $\chi^0_1$-MSSM-LSP.  For
$m_{\tilde{g}}=1\ {\rm TeV}$, $1.5\ {\rm TeV}$, and $2\ {\rm TeV}$,
$L^{\rm (eff)}$ and the cross section for the SUSY events are given by
$7.1\ {\rm m}$ and $2050\ {\rm fb}$, $8.1\ {\rm m}$ and $150\
{\rm fb}$, and $9.4\ {\rm m}$ and $20\ {\rm fb}$,
respectively.  As in the case of $\chi^0_1$-MSSM-LSP, we can see that
$L^{\rm (eff)}$ is insensitive to the mass spectrum of the
superparticles.  In Fig.\ \ref{fig:taumax_stau}, we show the lifetime
which gives $N_{\rm dec}=10$.  Again, when $\tau\lesssim
O(10^{-(3-5)}\ {\rm sec})$, we can expect the decay of the MSSM-LSP
inside the detector.

Comparing Figs.\ \ref{fig:taumax_bino} and \ref{fig:taumax_stau}, it
is understood that $N_{\rm dec}$ does not depend much on the mass
spectrum of the superparticles as far as the masses of the colored
superparticles are fixed.  This fact indicates that $N_{\rm dec}$ is
mostly determined by the total cross section for the SUSY events (as
well as by the lifetime of the MSSM-LSP).  (See Eq.\ (\ref{N_dec}).)

So far, we have seen that a sizable number of MSSM-LSPs decay inside
the detector when $\tau\lesssim 10^{-(3-5)}\ {\rm sec}$.  However,
this does not necessarily mean that the decay of the MSSM-LSP can be
easily observed.  In particular, in our calculation of $L^{\rm
  (eff)}$, we have assumed that the typical size of the fiducial
region is $\sim 10\ {\rm m}$.  (More accurately, see Eqs.\
(\ref{lmax}) and (\ref{boundary}).)  This implies that, in the case of
the ATLAS detector, the decay of the MSSM-LSP may have to be
identified by using only the last layer of the muon chamber.  Such an
analysis requires very high efficiency of detecting signals of the
decay in the muon detector.  In some cases, more conservative procedure
to confirm the decay may be necessary.  For example, for the case of
the $\chi^0_1$-MSSM-LSP, some possibilities are (i) to require the
decay of $\chi^0_1$ inside the inner tracker region so that we can
detect non-pointing photon using the electromagnetic calorimeter
\cite{Kawagoe:2003jv}, or (ii) to look for a decay in the calorimeters
to see the decay vertex.  For the case of $\tilde{\tau}$-MSSM-LSP, as
we have mentioned, we can require (i) no hits in the muon detector for
high $p_T$ charged particles observed by the inner detectors, or (ii)
the discovery of the decay vertex of $\tilde{\tau}$ in the
calorimeters or inner trackers.  Then, the size of the fiducial volume
to study the decay is reduced.  At the ATLAS detector, the inner
radius of the muon chamber, the outer radius of the hadron
calorimeter, and the outer radius of the inner tracker region are
about $5\ {\rm m}$, $4.25\ {\rm m}$, and $103\ {\rm cm}$, respectively
\cite{AtlasTdr}.  Importantly, $L^{\rm (eff)}$ given in Eq.\
(\ref{Leff}) is proportional to the (typical) length of the fiducial
region.  Thus, if the MSSM-LSP is required to decay in the
calorimeters or in the inner tracker, $L^{\rm (eff)}$ is expected to
be reduced by the factor of $2-10$, and hence the maximum value of the
lifetime with which sizable decay events are expected is.  Even with
such a smaller value of $L^{\rm (eff)}$, as Figs.\
\ref{fig:taumax_bino} and \ref{fig:taumax_stau} indicate, a
significant number of the MSSM-LSP may decay in the fiducial region in
the models we are interested in, as we see in the following.  The
understanding of the efficiency to find the signal of the decay should
require extensive studies of the detector effects, which is beyond the
scope of this letter.  Here, we simply assume that the signals of the
decay can be somehow identified in the following discussion.

Let us now consider implications of the search for the decay of the
long-lived MSSM-LSP.  First, we consider the gauge-mediated model.
When the Bino-like neutralino is the MSSM-LSP, it decays into gravitino
and a gauge boson (i.e., photon or $Z$-boson), and the lifetime of
$\chi^0_1$ is estimated as
\begin{eqnarray}
  \tau \simeq   2 \times 10^{-5}\ {\rm sec} \times 
  \left( \frac{m_{\chi^0_1}}{200\ {\rm GeV}} \right)^{-5}
  \left( \frac{m_{3/2}}{100\ {\rm keV}} \right)^{2},
\end{eqnarray}
where $m_{3/2}$ is the gravitino mass.  In the case where
$\tilde{\tau}$ is the MSSM-LSP, it decays into gravitino and the
tau-lepton, and the lifetime of $\tilde{\tau}$ is given by
\begin{eqnarray}
  \tau \simeq 
  2 \times 10^{-5}\ {\rm sec} \times 
  \left( \frac{m_{\tilde{\tau}}}{200\ {\rm GeV}} \right)^{-5}
  \left( \frac{m_{3/2}}{100\ {\rm keV}} \right)^{2}.
\end{eqnarray}
Thus, a sizable amount the MSSM-LSP decay is possible in both cases if
$m_{3/2}\lesssim O(100\ {\rm keV}-1\ {\rm MeV})$; in particular, using
the fact that too large $m_{\tilde{g}}$ is not preferred from the
naturalness point of view, the number of the decay events can be as
large as $O(100)$ in such a case.

Another important case is with $R$-parity violation.  In particular,
with the introduction of the following bi-linear $R$-parity breaking
interaction into the SUSY breaking terms:
\begin{eqnarray}
  {\cal L}_{\rm RPV} = B_{\rm RPV} \tilde{L} H_u + {\rm h.c.},
  \label{L_RPV}
\end{eqnarray}
with $\tilde{L}$ and $H_u$ being the slepton and the up-type Higgs
boson, respectively, it was pointed out that the EGRET and HEAT
anomalies can be simultaneously explained if the gravitino is the LSP
\cite{Ibarra:2008qg,Ishiwata:2008cu}.  (Here and hereafter, we neglect
the generation index for sleptons for simplicity.)  Such a scenario
works for any kind of the MSSM-LSP as far as the lifetime of the
gravitino is $\sim 10^{26}\ {\rm sec}$.  With the $R$-parity violating
interaction given in Eq.\ (\ref{L_RPV}), gravitino $\psi_\mu$
dominantly decays as $\psi_\mu\rightarrow\nu Z$ or $l^\pm W^\mp$.
When the gravitino is heavier than the weak bosons, the lifetime of
the gravitino is estimated as \cite{Ishiwata:2008cu}
\begin{eqnarray}
  \tau_{3/2} \simeq 7\times 10^{25}\ {\rm sec}
  \times
  \left( \frac{\kappa}{10^{-9}} \right)^{-2}
  \left( \frac{m_{3/2}}{200\ {\rm GeV}} \right)^{-3},
\end{eqnarray}
where $\kappa=B_{\rm RPV}/m_{\tilde{\nu}}^2$ (with $m_{\tilde{\nu}}$
being the sneutrino mass) is the VEV of the sneutrino in units of the
VEV of the standard-model like Higgs boson.  With the $R$-parity
violation given in Eq.\ (\ref{L_RPV}), the MSSM-LSP may also decay
dominantly via the $R$-parity violating interaction.  Then, when the
Bino-like neutralino is the MSSM-LSP, it decays into a neutrino and a
standard-model boson with the lifetime
\begin{eqnarray}
  \tau \simeq
  1 \times 10^{-6}\ {\rm sec} \times
  \left( \frac{\kappa}{10^{-9}} \right)^{-2}
  \left( \frac{m_{\chi^0_1}}{200\ {\rm GeV}} \right)^{-1}.
\end{eqnarray}
When $\tilde{\tau}$ is the MSSM-LSP, it decays into the tau lepton and
a neutrino, and the lifetime is given by
\begin{eqnarray}
  \tau \simeq
  3 \times 10^{-5}\ {\rm sec} \times
  \left( \frac{\kappa}{10^{-9}} \right)^{-2}
  \left( \frac{m_{\tilde{\tau}}}{200\ {\rm GeV}} \right)^{-1}
  \left( \frac{m_{\tilde{B}}}{300~{\rm GeV}} \right)^2,
\end{eqnarray}
where $m_{\tilde{B}}$ is the mass of the Bino-like neutralino.
Requiring $\tau_{3/2}\sim 10^{26}\ {\rm sec}$ to simultaneously
explain the EGRET and HEAT anomalies, $\kappa$ is required to be $\sim
10^{-9}$ for $m_{3/2}\sim 200\ {\rm GeV}$, resulting in $\tau\sim
O(10^{-6}\ {\rm sec})$ ($O(10^{-5}\ {\rm sec})$) when the MSSM-LSP is
Bino-like neutralino (stau).  From the study of $N_{\rm dec}$ given
above, we expect sizable amount of the decay of the MSSM-LSP inside
the detector with such a lifetime.  In particular, when masses of
superparticles are at most $\sim 1\ {\rm TeV}$ to solve the
naturalness problem, the number of the MSSM-LSP decay inside the
detector is expected to be $\sim 100-1000$.  Thus, the search for the
decay of the long-lived MSSM-LSP should give an important test of the
scenario to explain the origins of anomalous $\gamma$-ray and positron
fluxes.

So far, we have discussed the discovery of the decay of long-lived
MSSM-LSP.  Importantly, once the decay is found, we can also constrain
the lifetime of the MSSM-LSP.  This fact can be easily understood from
Eq.\ (\ref{N_dec}); once the total number of the produced MSSM-LSP and
the number of the decay in the detector are both determined, the
lifetime $\tau$ is constrained using Eq.\ (\ref{N_dec}).

A relatively good determination of the lifetime is expected in
particular when the MSSM-LSP is $\tilde{\tau}$ (or other charged
superparticles) even if the decay length is much longer than the size
of the detector \cite{Ambrosanio:2000ik}.\footnote
{If the decay length $c\tau$ of the MSSM-LSP is comparable or smaller
  than the size of the detector, decrease of the decay point may be
  experimentally observed.  Then, the lifetime is determined by using
  the distribution of the decay points when $c\tau\gtrsim \mbox{a
    few}\ {\rm cm}$ \cite{Asai:2008sk}.}
The tracking information about the long-lived charged particle will be
available if it travels transverse length longer than $O(10\ {\rm
  cm})$.  Thus, once large amounts of $\tilde{\tau}$ samples become
available, we expect that the distribution function $f$ can be
directly determined from the experimental data.  Since we consider the
case where $c\tau$ is much longer than the size of the detector, most
of the staus do not decay inside the detector, and are observed as
energetic charged particles.  Then, if enough amounts of
$\tilde{\tau}$s are identified with the measurement of their momenta,
we can determine $f(\eta,v)$.

In discussing the identification of long-lived $\tilde{\tau}$ tracks,
it should be noted that the stau may be confused with the muon in
particular when $\tilde{\tau}$ does not decay inside the detector.
This fact makes the determination of $N_{\rm tot}$ (or, more
accurately, the number of events in which both of $\tilde{\tau}$s do
not decay inside the detector) non-trivial.  One possibility to
distinguish $\tilde{\tau}$ from the muon is to use the time-of-flight
information; for this purpose, the transition radiation tracker and
the muon system are useful in the ATLAS detector.  If the velocity of
$\tilde{\tau}$ is small enough, it takes sizable time to reach
trackers, calorimeters, and muon chamber after the production.  If the
velocity information is combined with the momentum information,
$\tilde{\tau}$ with small enough velocity can be distinguished from
energetic muon.  In our study, we require that the velocity of at
least one of the two staus should be smaller than $v_{\rm max}$ so
that the event can be identified as a SUSY event; we assume that, with
the velocity measurement, $\tilde{\tau}$ with $v<v_{\rm max}$ can be
distinguished from high $p_T$ muon whose velocity is almost the speed
of light.  We adopt several values of $v_{\rm max}$ to see how the
results depend on $v_{\rm max}$.  On the contrary, if one of the staus
decays inside the detector, it provides a striking signal, as we have
discussed.

Based on the above argument, we assume that the SUSY events can be
identified if there exists at least one $\tilde{\tau}$ which escapes
from the detector with $v<v_{\rm max}$, or if one of $\tilde{\tau}$s
decays inside the detector.  (Since we consider the case $c\tau\gg
L^{\rm (eff)}$, we safely neglect the case where both of
$\tilde{\tau}$s decay inside the detector.)  Then, we expect that we
can experimentally count the number of events where one of the
$\tilde{\tau}$s decays inside the detector and the other escapes from
the detector with its velocity smaller than $v_{\rm max}$.  We denote
the number of such events as $N_{\rm dec}^{(v'<v_{\rm max})}$.
Importantly, $N_{\rm dec}^{(v'<v_{\rm max})}$ can be calculated as a
function of $\tau$ after the experimental determination of the
velocity distribution of $\tilde{\tau}$ with the above-mentioned type
of SUSY events.  Thus, with the measurement of $N_{\rm
  dec}^{(v'<v_{\rm max})}$, a determination of $\tau$ is possible.
Since we are interested in the case where $N_{\rm dec}^{(v'<v_{\rm
    max})}$ is much smaller than the total number of $\tilde{\tau}$s
observed, the statistical error in the determination of the lifetime
is dominantly from $N_{\rm dec}^{(v'<v_{\rm max})}$; $\delta\tau\simeq
\tau/\sqrt{N_{\rm dec}^{(v'<v_{\rm max})}}$.

If the correlation of the velocities of two staus is weak, $N_{\rm
  dec}^{(v'<v_{\rm max})}$ is given by
\begin{eqnarray}
  N_{\rm dec}^{(v'<v_{\rm max})} \simeq N_{\rm tot}^{(v<v_{\rm max})}
  \frac{L^{\rm (eff)}}{c\tau},
  \label{N_dec(v<vmax)}
\end{eqnarray}
where $N_{\rm tot}^{(v<v_{\rm max})}$ is the number of $\tilde{\tau}$s
which satisfy $v<v_{\rm max}$, which is also experimentally
measurable.  With our MC analysis, we have confirmed that Eq.\
(\ref{N_dec(v<vmax)}) holds with a good accuracy.  Thus, even though
more accurate relation between $N_{\rm dec}^{(v'<v_{\rm max})}$ and
$\tau$ can be obtained once the experimental data become available, we
use Eq.\ (\ref{N_dec(v<vmax)}) to estimate how well the lifetime can
be constrained for simplicity.

\begin{figure}[t]
 \centerline{\epsfxsize=0.55\textwidth\epsfbox{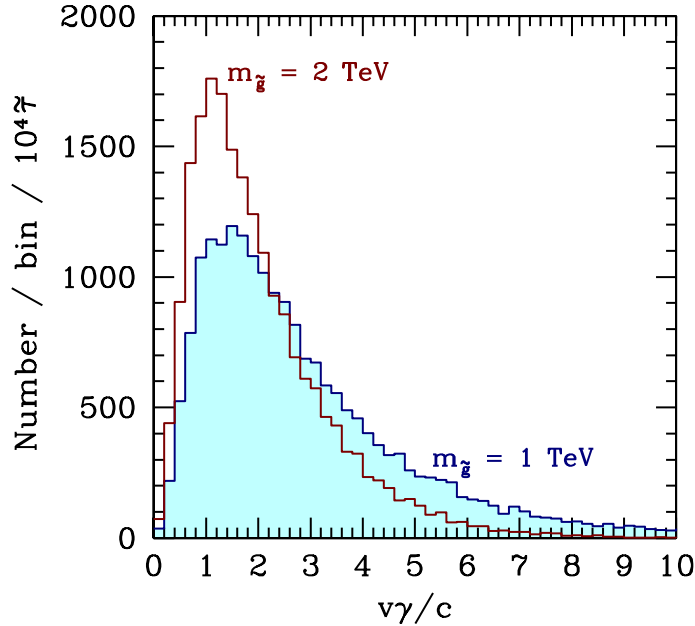}}
 \caption{\small Distribution of $v\gamma/c$ of the final-state stau
   with $10^4$ samples.  We take $m_{\tilde{g}}=1\ {\rm TeV}$ (shaded)
   and $2\ {\rm TeV}$ (unshaded).}
\label{fig:histbgstau}
\end{figure}

In order to see the effect of the velocity cut, we show the
distribution of $v\gamma$ of $\tilde{\tau}$.  The parameter $\Lambda$
is taken to be $66\ {\rm TeV}$ and $141\ {\rm TeV}$ so that the gluino
mass is given by $m_{\tilde{g}}=1\ {\rm TeV}$ and $2\ {\rm TeV}$,
respectively.  The result is shown in Fig.\ \ref{fig:histbgstau}.
From the figure, we can see that $\tilde{\tau}$ acquires large
velocity in average.  Thus, if one imposes relatively severe cut on
the velocity of $\tilde{\tau}$ to reduce the muon background,
significant amounts of the stau events will be also discarded.  (See
the following discussion.)

Next, we calculate $N_{\rm tot}^{(v<v_{\rm max})}$ for $v_{\rm
  max}=0.8c$ and $0.9c$; the lifetime which gives $N_{\rm
  tot}^{(v<v_{\rm max})}=10$ is also shown in Fig.\
\ref{fig:taumax_stau}.  (Notice that $N_{\rm dec}$ corresponds to
$N_{\rm dec}^{(v'<v_{\rm max})}$ with $v_{\rm max}=c$.)  Assuming that
all the decay events are identified, Fig.\ \ref{fig:taumax_stau} shows
contours on $m_{\tilde{g}}$ vs.\ $\tau$ plane on which the statistical
uncertainty in the determination of $\tau$ is about $30\ \%$ (i.e.,
$\delta\tau= \tau/\sqrt{10}$).  In order to determine $\tau$ at this
level, the lifetime is required to be, for $v_{\rm max}=c$ and $0.9c$,
for example, shorter than $1\times 10^{-3}$ and $4\times 10^{-4}\ {\rm
  sec}$ ($8\times 10^{-5}$ and $4\times 10^{-5}\ {\rm sec}$, $1\times
10^{-5}$ and $8\times 10^{-6}\ {\rm sec}$) for $m_{\tilde{g}}=1\ {\rm
  TeV}$ ($1.5\ {\rm TeV}$, $2\ {\rm TeV}$), respectively.  Notice
that, with a smaller input value of $\tau$, better determination of
the lifetime is expected.

As one can see, the sensitivity becomes worse as we require smaller
value of $v_{\rm max}$.  This is because most of the staus are
produced with relatively large velocity.  (See Fig.\
\ref{fig:histbgstau}.)  Importantly, the muon system of the ATLAS
detector is expected to have a good time resolution of about $0.7\
{\rm ns}$, and the velocity measurement is possible with the accuracy
of $\delta v/v^2 \simeq 0.03c^{-1}$ \cite{AtlasTdr}.  Thus, good
separation of $\tilde{\tau}$ from $\mu$ may be possible even with a
relatively large value of $v_{\rm max}$.  In addition, as we have
mentioned, the SUSY event may be identified even without imposing the
velocity cut if one of the staus decays inside the detector.  Thus, if
the total number of staus (with any velocity) can be somehow
determined, the velocity cut is unnecessary.  For example, if the
correlation of the velocities of two staus is experimentally checked
to be negligible, the total number of stau events can be calculated
from the number of events with $v_1<v_{\rm max}<v_2$ and that of
$v_1,v_2<v_{\rm max}$, where $v_1$ and $v_2$ are velocities of two
staus (with $v_1<v_2$).  In such a case, we can adopt the result with
$v_{\rm max}=c$.

Such a measurement of the lifetime provide a quantitative information
about underlying parameters.  For example, in the gauge-mediated
model, the measurement of the lifetime is possible when the gravitino
is lighter than $O(100\ {\rm keV}-1\ {\rm MeV})$, and consequently,
the gravitino mass can be determined assuming that $\tilde{\tau}$
decays into the gravitino.  Then, we can obtain an important
information about the SUSY breaking scale.\footnote
{For other possibilities, see \cite{GravAtColliders}.}
In addition, in the model with $R$-parity violation, the size of the
$R$-parity violating coupling constant can be determined.

Finally, we comment on the case where the lightest neutralino is the
MSSM-LSP.  Even though the discovery of the decay of (long-lived)
$\chi^0_1$, which is already very spectacular, may be possible, the
precise determination of $N_{\rm tot}$, which is necessary for the
determination of the lifetime of $\chi^0_1$, is very non-trivial.
This is because $\chi^0_1$ is neutral and we cannot directly count the
number of $\chi^0_1$s which do not decay inside the detector.  If the
total number of SUSY events can be somehow estimated, however, the
discovery of the decay of $\chi^0_1$ gives significant information
about the property of the MSSM-LSP.  Once the SUSY events are found,
we expect that the order-of-magnitude estimation of the number of SUSY
events can be performed, which gives a bound on $N_{\rm tot}$.  Then,
the measurement of $N_{\rm dec}$ gives some information about the
lifetime.

{\it Acknowledgments:}
This work was supported in part by Research Fellowships of the Japan
Society for the Promotion of Science for Young Scientists (K.I.), and
by the Grant-in-Aid for Scientific Research from the Ministry of
Education, Science, Sports, and Culture of Japan, No.\ 19540255
(T.M.).

\end{document}